\documentclass[twocolumn,preprintnumbers,amsmath,amssymb]{revtex4}

\usepackage{array}
\usepackage{graphicx}
\usepackage{tabularx}
\usepackage{longtable}
\usepackage{dcolumn} 
\usepackage{bm}

\begin{document}

\title{Structure-related transport properties of A-site ordered perovskite Sr$_{3}$ErMn$_{4-x}$Ga$_x$O$_{10.5-d}$
}

\author{W. Kobayashi$^{\dagger}$}
\address{Waseda Institute for Advanced Study, Waseda University, 
Tokyo 169-8050, Japan}

\author{T. Ishibashi}
\address{Department of Physics, Waseda University, 
Tokyo 169-8555, Japan}

\date{\today}

\begin{abstract}

We report x-ray diffraction, resistivity, thermopower, and magnetization of 
Sr$_{3}$ErMn$_{4-x}$Ga$_x$O$_{10.5-d}$, in which A-site ordered tetragonal phase appears above $x=1$, 
and reveal that the system exhibits typical properties seen in the antiferromagnetic insulator with Mn$^{3+}$. 
We succeed in preparing both A-site ordered and disordered phases for $x=1$ in 
different preparation conditions, and observe a significant decrease of the resistivity in the disordered phase. 
We discuss possible origins of the decrease focusing on the dimensionality and the disordered effect.

\end{abstract}

\maketitle

\section{Introduction}

Perovskite oxide is denoted as $AB$O$_3$, where $A$ and $B$ represent lanthanide (and/or alkaline earth 
elements) and transition-metal elements, respectively. $B$ ion is surrounded by six oxygen ions, and 
$B$O$_6$ octahedron is formed. The octahedron mainly contributes electrical and magnetic proerties 
of the perovskite oxide. According to the ionic radius of $A$ ion, the bond angle $\angle$ $B$-O-$B$ 
deviates from 180$^{\circ}$, which causes a change of the bandwidth. 
In a substituted system of $R^{3+}_{1-x}$$A^{2+}_{x}$$B$O$_3$, the carrier concentration 
is also controlled as well as the bandwidth. 

The A-site ordered manganese oxide $R$BaMn$_2$O$_6$ ($R:$ lanthanide) 
has attracted much attention because of significant physical phenomena 
such as the large magnetoresistance of 1,000 \% at room temperature \cite{nakajima0}, 
charge and orbital orderings at high temperatures \cite{nakajima1}. 
These properties attribute to 
A-site ordering working as "a periodic" Coulomb potential 
which stabilizes charge, spin, and/or orbital orderings of the electrons on B-site \cite{motome}. 
Owing to the randomness, A-site disordered phase of $R_{0.5}$Ba$_{0.5}$MnO$_3$ displays 
the spin-glass state or the itinerant ferromagnetic state instead of the charge ordered state seen in the ordered phase 
\cite{nakajima1,akahoshi}. 

The A-site ordered cobalt oxide Sr$_3$YCo$_4$O$_{10.5}$ also exhibits peculiar 
magnetic properties with a high ferromagnetic (ferrimagnetic) transition 
temperature of 335 K \cite{kobayashi2}, which is in contrast with a spin state crossover 
near 100 K in LaCoO$_3$ \cite{asai}. 
In this compound, the A-site ordering stabilizes oxygen deficient ordering, 
which makes a volume of the CoO$_6$ octahedron larger than that of LaCoO$_3$ \cite{ishiwata} and 
gives a different coordination number of Co$^{3+}$ from LaCoO$_3$. These modifications 
stabilize high-spin and/or intermediate-spin states of Co$^{3+}$ even 
at low temperatures, which causes the peculiar magnetic properties. 
Hence, partial substitution for the A-site or B-site cations strongly affects the spin state leading to significant suppression 
of the magnetic order; Ca substitution for Sr site \cite{yoshida} 
and 6\%-Mn doping in the B site \cite{kobayashi4} destroys the room-temperature ferromagnetism of 
Sr$_3$YCo$_4$O$_{10.5}$. 

Recently, a new A-site ordered perovskite Y$_{0.8}$Sr$_{2.2}$Mn$_2$GaO$_{7.9}$ 
(Sr$_{2.93}$Y$_{1.07}$ Mn$_{2.66}$Ga$_{1.34}$O$_{10.53}$) was reported 
by Gillie {\it et al.} \cite{gillie}, which is isostructural to Sr$_{3}$YCo$_4$O$_{10.5}$ \cite{istomin1,withers}. 
As shown in Fig. 1, this compound has an octahedral site where Mn$^{3+}$ ions mainly occupy 
with about 10\%-Ga ions intermixed and a tetrahedral site where both Mn$^{3+}$ and Ga$^{3+}$ ions occupy. 
They found an antiferromagnetic state below 100 K in this material showing 
that Y$_{0.8}$Sr$_{2.2}$Mn$_2$GaO$_{7.9}$ was a antiferromagnetic insulator, 
however, they did not report the transport properties. 
We have prepared polycrystalline samples of Sr$_{3}$ErMn$_{4-x}$Ga$_x$O$_{10.5-d}$ ($x=$ 0, 0.5, 1, 2) 
where $d$ represents oxygen deficiency 
and investigated the transport and magnetic properties in relation to the strcture. 
We have succeeded in preparing both A-site ordered and disordered phases for $x=1$ using 
different preparation conditions and observe a significant decrease of the resistivity in the disordered phase. 
We attribute this to a change of dimensionality in conduction and/or A-site disordered effect. 
This material can be a good playground to study order-disorder effect on the electronic states of the 
antiferromagnetic insulator with Mn$^{3+}$. 

\begin{figure}[t]
\begin{center}
\vspace*{0cm}
\includegraphics[width=4cm,clip]{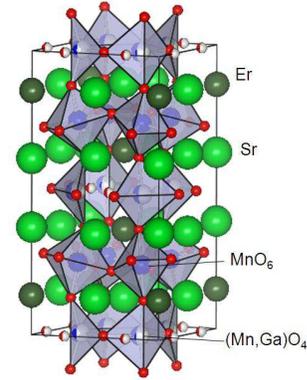}
\caption{(Color online) Crystal structure of Sr$_{3}$ErMn$_{4-x}$Ga$_x$O$_{10.5-d}$ ($x\geq 1$). 
}
\end{center}
\end{figure}

\section{Experiments}

Polycrystalline samples of Sr$_{3}$ErMn$_{4-x}$Ga$_x$O$_{10.5-d}$ ($x=0, 0.5, 1$, and 2)
were prepared by a solid state reaction. Stoichiometric amounts of 
SrCO$_3$, Er$_2$O$_3$, Mn$_3$O$_4$, and Ga$_2$O$_3$ were mixed, 
and the mixture was sintered at 1250 $^{\circ}$C for 12 h for $x=0$ and 0.5, 6 h for $x=1$ and 2 
in N$_2$ flow ($100-200$ ml/min). 
Then, the product was finely ground, pressed into a pellet, and sintered at 1250 $^{\circ}$C for 12 h for $x=0$, 
24 h for $x=0.5$, and 10 h for $x=1$ and 2 in N$_2$ flow ($100-200$ ml/min). 
The second process was repeated 2 times for $x=0.5$, and once for $x=1$ and 2 
with intermediate grindings and pelletizings. 
As shown in Fig. 2(c), $x=1$ sample exhibits A-site ordered structure. We sintered the $x=1$ sample 
again using the second process, and found the sample shows a disordered structure shown in Fig. 3.

The x-ray diffraction of the sample was measured using a standard diffractometer with Cu K$\alpha$ 
radiation as an x-ray source in the $\theta -2\theta $ scan mode. 
The structural simulations were performed using a RIETAN-2000 program \cite{izumi}. 
The resistivity was measured by a four-probe method in a liquid He cryostat. 
The thermopower was measured using a steady-state technique in a liquid He cryostat 
with copper-constantan thermocouple to detect a small temperature 
gradient of about 1 K/cm. The magnetization was measured from 5 to 400 K by a 
commercial superconducting quantum interference device (SQUID, Quantum Design MPMS). 

\section{Results and discussion}

\begin{figure}[t]
\begin{center}
\vspace*{0cm}
\includegraphics[width=6cm,clip]{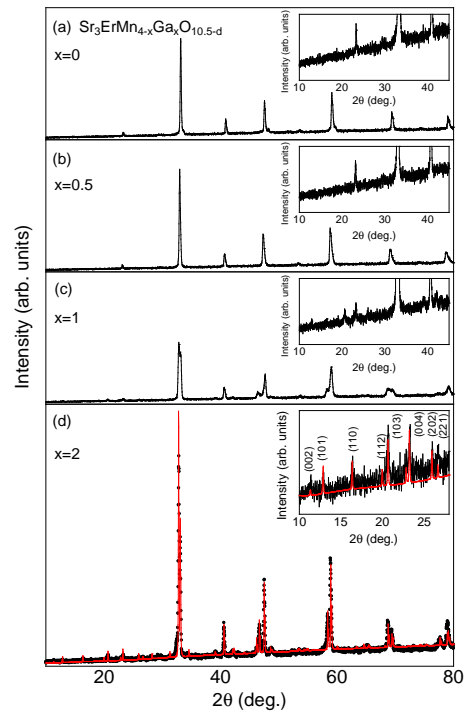}
\caption{(Color online) (a)-(d) X-ray diffraction patterns of Sr$_{3}$ErMn$_{4-x}$Ga$_x$O$_{10.5-d}$ ($x=0, 0.5, 1$, and 2). 
The insets represent the magnified x-ray patterns at low angles. 
The red line of the inset of Fig. 2(d) represents a simulated pattern.  
}
\end{center}
\end{figure}

Figure 2 shows the x-ray diffraction patterns of Sr$_{3}$ErMn$_{4-x}$Ga$_x$O$_{10.5-d}$ ($x=0, 0.5, 1$, and 2). 
All the peaks are indexed as a cubic cell of the space group $Pm3m$ with the lattice parameter of 
$a\sim $ 3.8 and 3.85 \AA~for $x=$ 0 and 0.5, respectively. This cubic cell is also seen in 
Sr$_{1-x}$Y$_{x}$CoO$_{3-\delta }$ with small $x$ \cite{kobayashi2}. 
With increasing Ga content $x$, crystal structure changes from the cubic perovskite 
to a tetragonal A-site ordered perovskite (space group: $I4/mmm$, Fig. 1) with the lattice parameter of 
$a\sim $ 7.63 and 7.65 \AA,~and $c\sim $ 15.58 and 15.56 \AA~for $x=$ 1 and 2, respectively. 
This result is consistent with the structural analysis by Gillie {\it et al.} \cite{gillie}. 
As shown in the inset of Fig. 2(d), superstructure peaks corresponding to the A-site ordering are observed, 
while they do not appear for $x=$ 0 and 0.5 samples. 
Ga ions selectively occupy the tetrahedral site to stabilize the A-site ordered structure as shown in Fig. 1 \cite{gillie}. 
Thus, $x=1$ is a minimal amount to stabilize the structure. 
Gillie {\it et al.} reported that the oxygen content of Sr$_{2}$YMn$_{2}$GaO$_{7.9}$ is 
7.9 showing formal valence of Mn ion is almost 3+. 
Thus, it is assumed that the formal valence of Mn ion is also 3+ in the ordered compounds presented here.

Figure 3 shows the x-ray diffraction patterns of Sr$_{3}$ErMn$_{3}$GaO$_{10.5-d}$ 
with different preparation conditions. Obviously, two patterns are different; 
one sample was identified to the tetragonal ordered phase (hereafter this is denoted by O sample), 
and the other was identified to the disordered-cubic perovskite phase (D sample). 
Similar structures are originally found in 
$R$BaMn$_{2}$O$_{6}$/$R_{0.5}$Ba$_{0.5}$MnO$_3$ systems \cite{nakajima1}. 
We would like to say that the O sample is metastable so that the longer sintering stabilizes the disordered phase. 
Thus, this composition is just on the verge of order and disorder. 
As shown in the inset of Fig. 3, the superstructure peaks of D sample are hardly visible. 

\begin{figure}[t]
\begin{center}
\vspace*{0cm}
\includegraphics[width=6.5cm,clip]{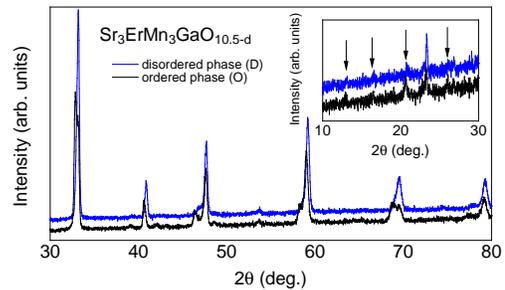}
\caption{(Color online) X-ray diffraction patterns of ordered and disordered Sr$_{3}$ErMn$_{2}$Ga$_2$O$_{10.5-d}$. 
The inset represents the magnified x-ray pattern at low angles. 
}
\end{center}
\end{figure}

Figure 4 (a) shows the thermopower of Sr$_{3}$ErMn$_{4-x}$Ga$_x$O$_{10.5-d}$ 
($x=0, 0.5, 1$, and 2). The magnitude of the thermopower is 
between $60$ and $110$ $\mu $V/K, and the sign is negative showing that the carriers are electrons. 
Assuming that the formal Mn valence is 3+, we expect that a tiny amount of electrons on 
Mn$^{2+}$ moves in the background of Mn$^{3+}$ as 
shown in the inset of Fig. 4(a). 
Using an extended Heikes formula \cite{koshibae}, 
the valence of Mn ion was evaluated to be 2.71+ at 300 K corresponding to $d=0.43$ 
for $x=1$ sample with spin 
degeneracy term of $g_{\rm Mn^{2+}}=$6 and $g_{\rm Mn^{3+}}=$5.  

Figure 4 (b) shows the resistivity of Sr$_{3}$ErMn$_{4-x}$Ga$_x$O$_{10.5-d}$ 
($x=0, 0.5, 1$, and 2). Semiconducting temperature dependence 
is observed for all the samples. With $x$, the magnitude of the resistivity decreases mainly due to 
decrease of scattering centers of the Ga ions. As seen in the inset of Fig. 4(b), the temperature 
dependence is described by an activation-type conduction 
$\rho =\rho _0$exp($\frac{E_{\rm g}}{k_{\rm B}T}$) where $E_{\rm g}$ represents activation energy above 200 K. 
The activation energy $E_{\rm g}$ was evaluated to be 0.133, 0.146, 0.149, and 0.246 eV for 
$x=$0, 0.5, 1, and 2, respectively. 

\begin{figure}[t]
\begin{center}
\vspace*{0cm}
\includegraphics[width=6cm,clip]{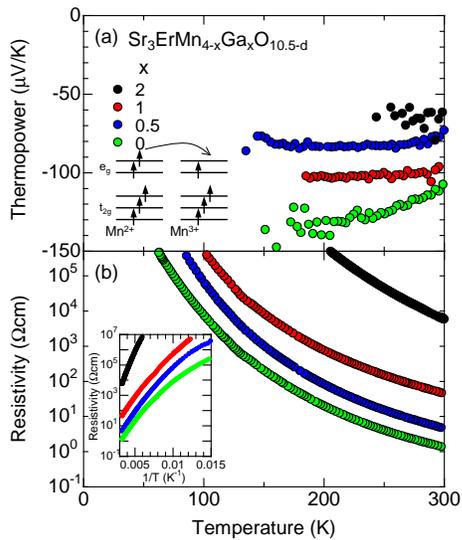}
\caption{(Color online) (a) Thermopower and (b) resistivity of Sr$_{3}$ErMn$_{4-x}$Ga$_x$O$_{10.5-d}$ 
($x=$0, 0.5, 1, and 2). 
}
\end{center}
\end{figure}

Figure 5(a) shows the magnetization of Sr$_{3}$ErMn$_{4-x}$Ga$_x$O$_{10.5-d}$ 
($x=0, 0.5, 1$, and 2). As shown in Figs. 5(b) and (c), the data was fitted by the Curie-Weiss law described by 
$\chi =\frac{C}{T-\theta _{\rm W}}+\chi _0$, where $C$, $\theta _{\rm W}$, and $\chi _0$ 
represent Curie constant, Weiss temperature and temperature independent term of the magnetic susceptibility, 
respectively. $C$ was evaluated to be 0.023, 0.019, 0.026, and 0.018 emu/K$\cdot $g for $x=0, 0.5, 1$, and 2, 
corresponding to 12.34, 11.14, 13.17, and 11.07 $\mu _{\rm B}$/f.u., respectively. 
These values are roughly explained by coexistence of 9.6 $\mu _{\rm B}$ 
of Er$^{3+}$ with $g=\frac{6}{5}$ and $J=\frac{15}{2}$ and 3.87 $\mu _{\rm B}$ and 4.90 $\mu _{\rm B}$ 
in the high-spin states of Mn$^{4+}$ and Mn$^{3+}$. 
$\chi _0$ was evaluated to be 1.39$\times$10$^{-5}$, 1.57$\times$10$^{-5}$, 0, and 0 emu/g for 
$x=0, 0.5, 1$, and 2 samples, respectively. Since $x=0$ and 0.5 samples show relatively good 
electric conductance compared with those of $x=1$ and 2 samples, 
this contribution may come from conducting electrons. 
All the samples exhibit negative $\theta _{\rm W}$ 
($-$45, $-$29, $-$52, and $-$30 K for $x=0, 0.5, 1$, and 2, respectively) 
implying antiferromagnetic interaction in this system. 
At around 60 K, the slope of $\chi ^{-1}$ changes showing an existence of magnetic anomaly for all the samples. 
Since the antiferromagnetism was observed in Sr$_2$YMn$_2$GaO$_{8-d}$ below 100 K \cite{gillie}, 
the anomaly at around 60 K in Sr$_{3}$ErMn$_{4-x}$Ga$_x$O$_{10.5-d}$ 
can be also related to antiferromagnetic order of Mn$^{3+}$. 

\begin{figure}[t]
\begin{center}
\vspace*{0cm}
\includegraphics[width=6cm,clip]{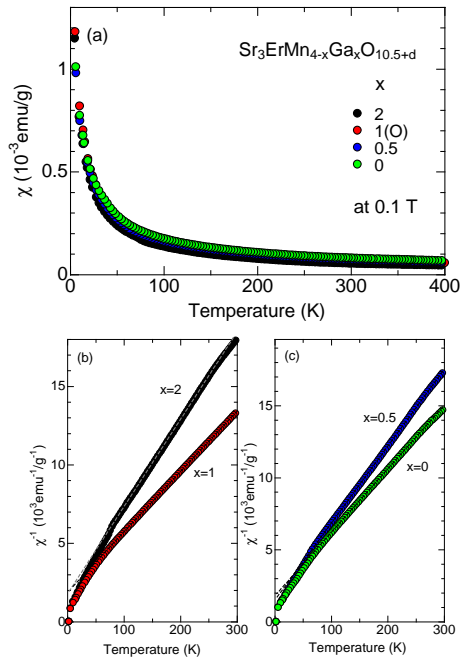}
\caption{(Color online) Magnetic susceptibility of Sr$_{3}$ErMn$_{4-x}$Ga$_x$O$_{10.5-d}$ 
($x=$0, 0.5, 1, and 2). The inset shows inverse susceptibility. 
}
\end{center}
\end{figure}

\begin{figure}[t]
\begin{center}
\vspace*{0cm}
\includegraphics[width=7cm,clip]{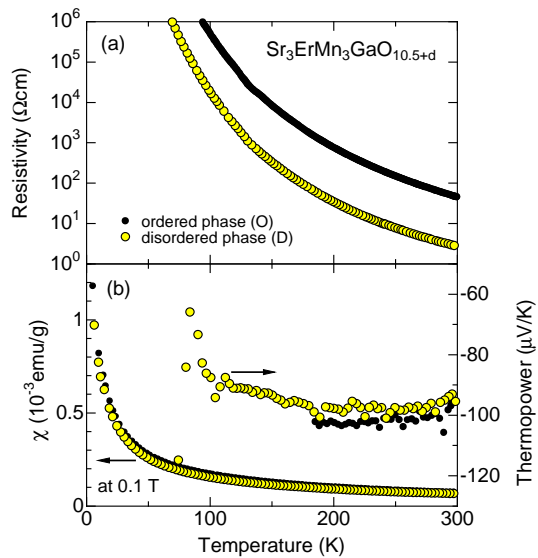}
\caption{(Color online) (a) Resistivity and (b) thermopower and magnetization of the ordered and disordered phases.  
}
\end{center}
\end{figure}

Lastly, we will discuss a difference of the transport and magnetic properties between 
the ordered (O sample) and disordered samples (D sample). 
Figure 6 (a) shows the temperature dependences of the resistivity of the O and D samples. 
The magnitude of the resistivity of the D sample is one order of magnitude smaller than that of the O sample, 
while the thermopower and magnetization of the D sample quite resemble those 
of the O sample as seen in Fig. 6(b). 
This strongly contrasts with the difference between the ferromagnetic-metal state of 
the disordered $R_{0.5}$Ba$_{0.5}$MnO$_3$ and charge-ordered 
insulating state of the ordered $R$BaMn$_2$O$_6$ \cite{nakajima1}. 
This variety of the properties comes from Mn$^{3.5+}$ with the charge degree of freedom, 
while Sr$_{3}$ErMn$_{4-x}$Ga$_x$O$_{10.5-d}$ has 
Mn$^{3+}$ without the charge degree of freedom causing the small difference between the properties.  

The difference of the resistivity shown in Fig. 6(a) is explained by several possible origins as follows: 
(1) a decrease of the carrier concentration, (2) an increase of the scattering time, 
(3) a change of dimensionality in conduction and (4) a change of electronic structure induced by the disordering of A site. 
As seen in Fig. 6(b), two samples exhibit almost the same magnitude of the thermopower of $-100$ $\mu$V/K. 
Since thermopower is a function of carrier concentration, the result shows 
that carrier concentration does not differ so much in the two samples. 
In addition, since the content $x$ of Ga ions which can be scattering centers is 1 in both samples, 
a possibility of (2) can be also denied. 
As stated in the introduction, Ga ion selectively occupies tetrahedral sites in the O phase, while 
Ga and Mn ions randomly occupies in the D phase. Thus, a dimensionality in conduction can change from 2D in the 
O sample  to 3D in the D sample, which can be a possible origin of the decrease of the resistivity. 
Another possibility is a change of electronic structure induced by the disordering of A site. 
A-site disordering generally induces a random potential in a perovskite oxide, which 
may affect the electronic structure of the material. 
Indeed, it is theoretically found that the introduced random potential in the Hubbard Hamiltonian 
causes an antiferromagnetic metallic phase instead of the antiferromagnetic insulating state \cite{shinaoka}. 
Although the origin of the difference of the resistivity is not clear at present, 
this system can be a good playground for investigating A-site disorder effect on antiferromagnetic 
insulator with Mn$^{3+}$. 

\section{summary} 

In summary, we have measured x-ray diffraction, resistivity, thermopower, and magnetization 
of Sr$_{3}$ErMn$_{4-x}$Ga$_x$O$_{10.5-d}$ system, 
in which A-site ordered tetragonal phase appears above $x=1$, and observed 
large negative thermopower, semiconducting conduction, and magnetic susceptibility with a kink at 
60 K implying antiferromagnetism. 
We succeed to prepare both A-site ordered and disordered phases for $x=1$ sample 
and observe a significant decrease of the resistivity in the disordered phase. 
We attribute this to a change of dimensionality in conduction and/or a change of electronic structure induced 
by the disordering of A sites. 

\section{acknowledgements}

We acknowledge I. Terasaki for fruitful discussion. This study was supported by the program entitled "Promotion of
Environmental Improvement for Independence of Young Researchers" under the
Special Coordination Funds for Promoting Science and Technology provided by
MEXT, Japan.


\begin{thebibliography}{99}

\bibitem[$^\dagger$]{Email1}email address: kobayashi-wataru@suou.waseda.jp

\bibitem{nakajima0}
	T. Nakajima and Y. Ueda, J. Appl. Phys. {\bf 98} (2005) 46108.  

\bibitem{nakajima1}
	T. Nakajima, H. Kageyama, H. Yoshizawa and Y. Ueda, 
	J. Phys. Soc. Jpn. {\bf 71} (2002) 2843. 

\bibitem{motome}
	Y. Motome, N. Furukawa, and N. Nagaosa, 
	Phys. Rev. Lett. {\bf 91} (2003) 167204. 

\bibitem{akahoshi}
	D. Akahoshi, M. Uchida, Y. Tomioka, T. Arima, Y. Matsui, and Y. Tokura, 
	Phys. Rev. Lett. {\bf 90} (2003) 177203. 

\bibitem{kobayashi2}
	W. Kobayashi, S. Ishiwata, I. Terasaki, M. Takano, I. Grigoraviciute, H. Yamauchi, and M. Karppinen, 
	Phys. Rev. B {\bf 72} (2005) 104408. 

\bibitem{asai}
	K. Asai, O. Yokokura, N. Nishimori, H. Chou, J. M. Tranquada, G. Shirane, S. Higuchi, Y. Okajima, and K. Kohn, 
	J. Phys. Soc. Jpn {\bf 50} (1994) 3025. 

\bibitem{ishiwata}
	S. Ishiwata, W. Kobayashi, I. Terasaki, K. Kato, and M. Takata, 
	Phys. Rev. B {\bf 75} (2007) 220406. 

\bibitem{yoshida}
	S. Yoshida, W. Kobayashi, T. Nakano, I. Terasaki, K. Matsubayashi, Y. Uwatoko, I. Grigoraviciute, 
	M. Karppinen, and H. Yamauchi, (submitted). 

\bibitem{kobayashi4}
	W. Kobayashi, S. Yoshida, and I. Terasaki, 
	Prog. Solid State Chem. {\bf 35} (2007) 355. 

\bibitem{gillie}
	L. J. Gillie, H. M. Palmer, A. J. Wright, J. Hadermann, G. Van Tendeloo, and C. Greaves, 
	J. Phys. Chem. Solids {\bf 65} (2004) 87. 

\bibitem{istomin1}
	S. Ya. Istomin, J. Grins, G. Svensson, O. A. Drozhzhin, V. L. Kozhevnikov, E. V. Antipov, and J. P. Attfield, 
	Chem. Mater. {\bf 15} (2003) 4012 

\bibitem{withers}
	R. L. Withers, M. James, and D. J. Goosens, 
	J. Solid State. Chem. {\bf 174} (2003) 198.

\bibitem{izumi}
	F. Izumi and T. Ikeda, 
	Mater. Sci. Forum {\bf 198} (2000) 321.

\bibitem{koshibae}
	W. Koshibae, K. Tsutsui, and S. Maekawa, 
	Phys. Rev. B {\bf 62} (2000) 6869. 

\bibitem{shinaoka}
	H. Shinaoka, and M. Imada, 
	Phys. Rev. Lett. {\bf 102} (2009) 016404. 

\end{thebibliography}
\end{document}